\documentclass[12pt]{iopart}
\usepackage{graphicx}
\usepackage{dcolumn}
\usepackage{bm}

\begin{document}

\title{Modeling Tunneling for the Unconventional Superconducting Proximity Effect}

\author{Parisa Zareapour,$^{1}$ Jianwei Xu,$^{1}$ Shu Yang F. Zhao,$^{1}$ Achint Jain,$^{1}$ Zhijun Xu,$^{2}$ T. S. Liu,$^{2,3}$ G.D. Gu,$^{2}$\ and Kenneth S. Burch $^{1,4}$}
\address{$^{1}$ Department of Physics and Institute for Optical Sciences, University of Toronto, 60 St George Street, Toronto, Ontario, Canada M5S 1A7.}%
\address{$^{2}$ Department of Condensed Matter Physics and Materials Science (CMPMS), Brookhaven National Laboratory, Upton, New York 11973, USA.}%
\address{$^{3}$ School of Chemical Engineering and Environment, North University of China, China.}%
\address{$^{4}$ Department of Physics, Boston College, 140 Commonwealth Avenue, Chestnut Hill, MA 02467.}%

\begin{abstract}

Recently there has been reinvigorated interest in the superconducting proximity effect, driven by predictions of the emergence of Majorana fermions. To help guide this search, we have developed a phenomenological model for the tunneling spectra in anisotropic superconductor-normal metal proximity devices. We combine successful approaches used in s-wave proximity and standard d-wave tunneling to reproduce tunneling spectra in d-wave proximity devices, and clarify the origin of various features. Different variations of the pair potential are considered, resulting from the proximity-induced superconductivity. Furthermore, the effective pair potential felt by the quasiparticles is momentum-dependent in contrast to s-wave superconductors. The probabilities of reflection and transmission are calculated by solving the Bogoliubov equations. Our results are consistent with experimental observations of the unconventional proximity effect and provide important experimental parameters such as the size and length scale of the proximity induced gap, as well as the conditions needed to observe the reduced and full superconducting gaps. 

\end{abstract}
\pacs{74.45.+c,74.72.-h,73.40.Gk}

\submitto{\SUST}
\maketitle

\section{Introduction}
In the proximity effect, superconductivity can be produced locally in a normal material placed in close contact with a superconductor\cite{PCvanSon:2011wx,Wolf:1982uv,Kastalsky:1991vh,Nguyen:1992tb}. Proximity-induced superconductivity in semiconductors has been gaining increasing attention lately\cite{2015NatCo...6E7130K, Cho:2013ij, PhysRevB.93.035307, Yang:2012eo, Koren:2011jc, Yang:2012co, Wang:2012if, Zhang:2011bl,2013ApPhL.103o2601L,2014PhRvL.112u7001X,Zareapour2014PRLsub}, due to the potential to generate new topologically protected\cite{Fu:2008gu,Beenakker:2011tp,Linder:2010hv} excitations and potential practical applications\cite{Asano:2009hi,Suemune:2006du,Hanamura:2002gk,2011PhRvL.107o7403S,2010PhRvL.104o6802R}. A few groups have turned their attention to the high critical temperature (T$_{c}$) superconducting proximity effect as it may generate novel pairing symmetries,\cite{PhysRevB.88.104514} as well as its technical advantages\cite{Lucignano:2012wj,Sato:2010cc,Tsuei:2013tf,Takei:2013dp,Hayat:2012jo}. Indeed beyond the large gaps and T$_{c}$, the cuprates have a small density of states and linear relation between energy and momentum, making it easier to match boundary conditions with topological insulators.\cite{Pannetier:2000wr,1982PhRvB..25.4515B}. Furthermore, the spatial extent of the proximity effect is set by either the coherence length or the ratio of the density of states and diffusion coefficients of the two materials.\cite{1991JAP....69.4137D,2013PhRvB..87m4511N} Thus the small density of states and the nearly two-dimensional electronic structure of the high T$_{c}$ superconductors suggests these systems are ready-made for proximity with topological insulators.

Indeed, numerous groups have demonstrated a high T$_{c}$ proximity effect with insulating cuprates\cite{2000PhRvL..85.3708D,1996PhyC..266..253D,Tarutani:1991cr,2004PhRvL..93o7002B,Kabasawa:vj}, Au quantum dots\cite{Sharoni:2004ba}, and in LCMO/YBCO thin films\cite{KorenProximityYBCO,weiManganiteYBCO,PhysRevB.67.214511,GolodPRB2013}. Recently we were able to achieve a proximity effect between Bi$_{2}$Sr$_{2}$CaCu$_{2}$O$_{8+\delta}$\cite{2012NatCo...3E1056Z} and two heavily doped topological insulators (Bi$_2$Se$_3$, Bi$_2$Te$_3$), which are known to superconduct with heavy doping\cite{Wray:2010uq,PhysRevB.85.125111} or pressure\cite{Zhang04012011}. This observation was enabled by an interpretation of the tunneling spectra, consistent with previous work on s-wave proximity junctions in superconductors. However these results are now controversial since ARPES experiments on thin film heterostructures conflict about the observation of proximity.\cite{Bi2212Bi2Se3FilmArpes,yilmaz2014absence,PhysRevB.90.085128} Part of the difficulty lies in the different means by which the junctions were fabricated. In addition there is a need for a framework to model the tunneling spectra of anisotropic superconductors when the proximity effect is present. Recently an approach to junctions with nanoscale disorder at the surface of a d-wave superconductor has emerged.\cite{ZhitlukhinaNRL16}  Here we employ a phenomenological approach that combines one used successfully in s-wave proximity devices and the standard approach for calculating tunneling into anisotropic superconductors. Nonetheless, we note there is a need for a more general theory that self-consistently calculates the induced superconductivity.

\begin{figure}
\includegraphics[width=0.48\textwidth]{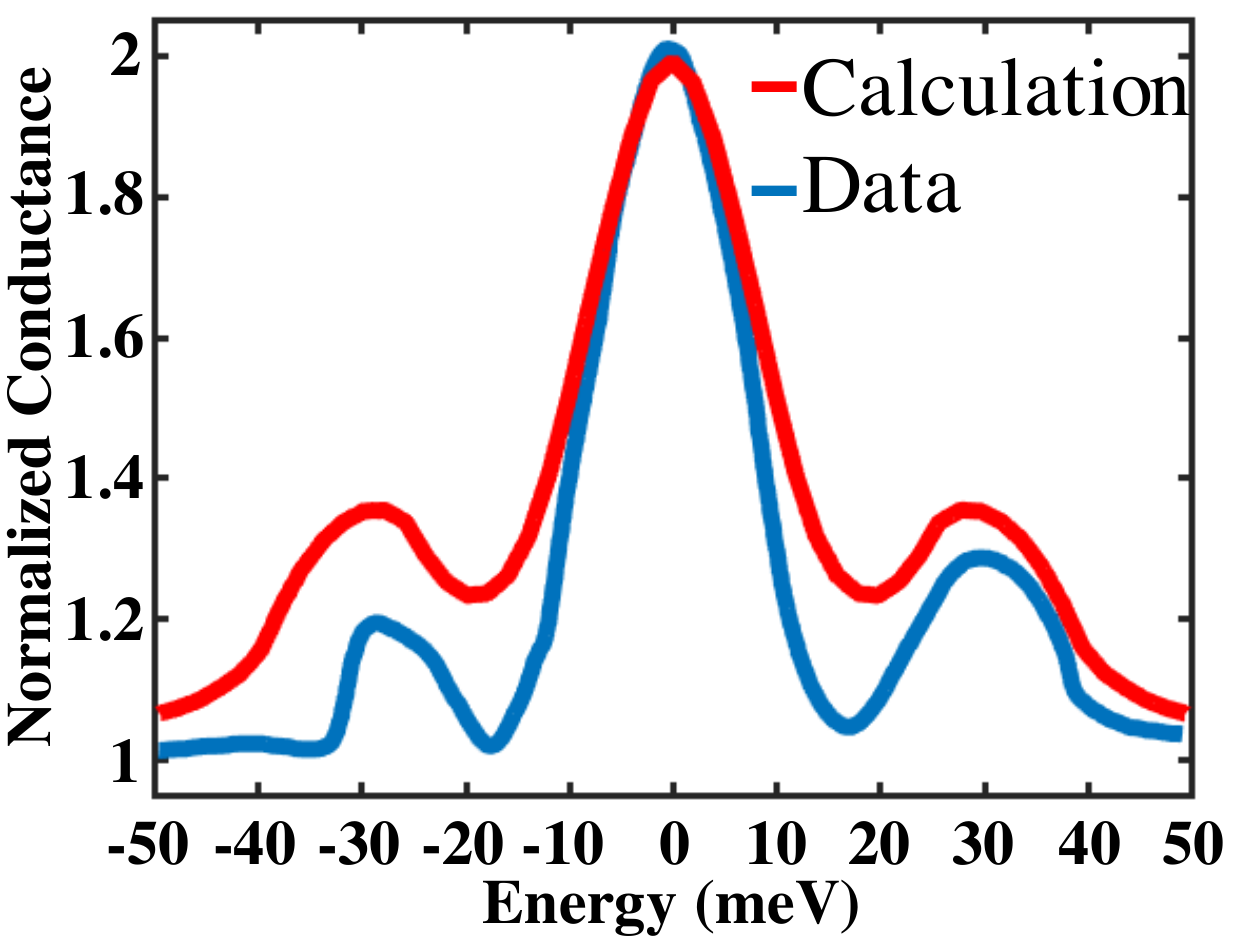}
\caption{\label{fig:BS} (color online) Differential conductance normalized to the normal state from an optimally-doped Bi$_2$Sr$_2$CaCu$_2$O$_{8+x}$/Bi$_2$Se$_3$ proximity device (red curve).\cite{2012NatCo...3E1056Z} The result of our model with parameters chosen match the data are shown in blue, parameters: $\Delta_{induced} = 10 mV$, $\Delta_{reduced} = 24 mV$, $\Delta_{0} = 40 mV$,  Z= 0.5, x$_S$ = 2.6$\epsilon_0$, x$_N$ = 1.3$\epsilon_0$.}
\end{figure}

Blonder, Tinkham, and Klapwijk (BTK)\cite{1982PhRvB..25.4515B} showed that the Bogoliubov equations are very suitable to describe the reflection and transmission of quasiparticles at a normal material/(s-wave) superconductor interface. They assumed that, at the interface, the superconducting gap ($\Delta(x)$) increases instantaneously from zero to a constant value in the superconductor ($\Delta_0$). The solutions of the Bogoliubov equations in the superconductor and the normal material are found by applying the appropriate boundary conditions. Ultimately the conductance is calculated using these wavefunctions to determine the transmission and reflection probabilities. Kashiwaya et al. later modified the BTK theory to take into account the momentum ($\vec{k}$) dependence of the superconducting order parameter often found in unconventional superconductors\cite{Kashiwaya:1996wz}. Here we extend the calculations of Kashiwaya et al. to study the effect of a gradual variation of the magnitude of the pair potential ($\Delta(x)$) near the superconducting-normal interface on the reflection and transmission coefficients of the quasiparticles. Ideally, $\Delta(x)$ should be calculated self-consistently, including the various possible superconducting symmetries that may arise on the normal material side. Interesting the exact form of superconductivity expected from the proximity effect with topological insulators remains controversial, especially for a d-wave superconductor.\cite{2013PhRvB..87v0506B,Haim:2016um} Thus we have chosen to focus on the case where the nature of the order parameter doesn't change across the interface. We note that our approach could applied to other combinations, by changing the form of the angular dependence of $\Delta(x)$, however to better explore the role of various parameters, we limit ourselves to a phenomenlogical $\Delta(x)$ that has worked well for s-wave proximity junctions.\cite{PCvanSon:2011wx} We find good agreement with existing differential conductance data obtained from an optimally-doped Bi$_2$Sr$_2$CaCu$_2$O$_{8+x}$/Bi$_2$Se$_3$ proximity device (figure \ref{fig:BS}). The resulting calculation demonstrates an induced and a reduced superconducting gap due to the proximity effect, consistent with previous observations of induced superconductivity via proximity to unconventional superconductors\cite{2012NatCo...3E1056Z}. Furthermore our model of $\Delta(x)$ allows us to vary different parameters to see their connection to the spectra observed, enabling us to reveal the origin of various features in the spectra. However we note that this model assumes the normal material already has a pairing potential that would enable the proximity effect, and that the induced superconductivity retains the same momentum dependence as the original superconductor. 

\begin{figure}
\includegraphics[]{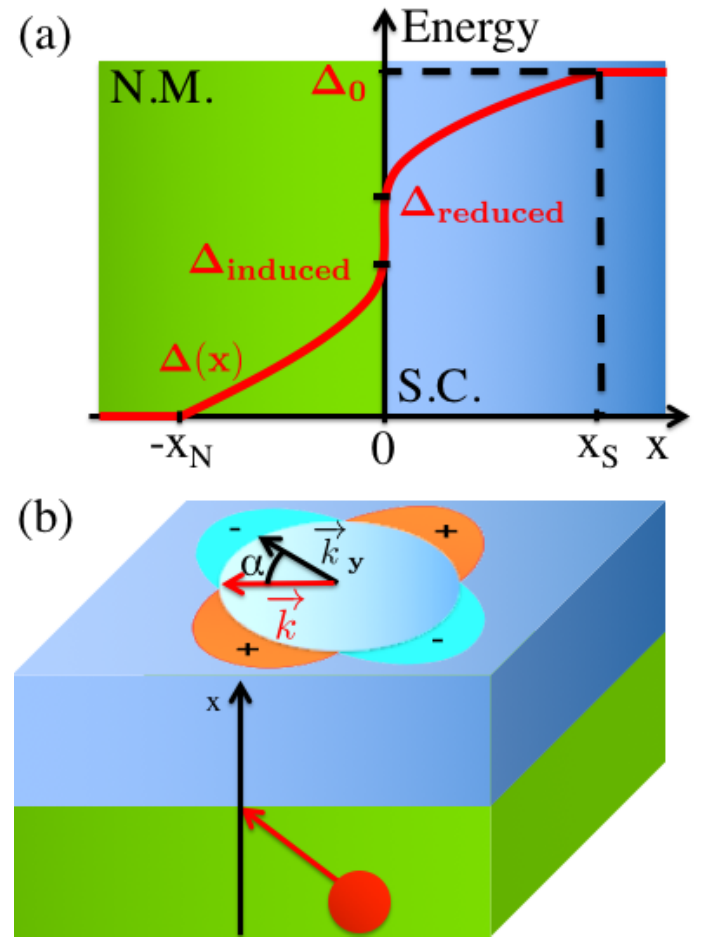}
\caption{\label{fig:gap} (color online) (a) Position dependence of the pair potential in a proximity junction, with the gap as a function of position determined by its value in the normal material at the interface ($\Delta_{induced}$), its value in the superconductor at the interface ($\Delta_{reduced}$) and in the bulk ($\Delta_{0}$) as well as the extent of the reduced ($x_{S}$) and induced ($x_{N}$). (b) The interface for scattering with the normal indicated (x-direction) as well as the in-plane d-wave gap in the superconductor. The peak in the gap is along the y, z axis and the direction of the scattering electron ($\vec{k}$) and its angle with the y-axis ($\alpha$) are shown. }
\end{figure}

At the interface between a superconductor and a normal material, the Cooper pairs can leak into the normal material and create a smaller superconducting gap (induced gap $\Delta_i$) locally. As shown in figure \ref{fig:gap}, this also results in the superconducting gap at the surface being suppressed (reduced gap $\Delta_r$). Some discontinuity in the gap function is allowed at the interface due to the presence of a barrier between the two materials. To simulate a realistic heterostructure, a delta function barrier potential of the form H$\delta$(x) is assumed at the interface (as is common to do in tunneling calculations\cite{1982PhRvB..25.4515B,Kashiwaya:1996wz}). To simplify the formulas, we define the dimensionless parameter, $Z(\Theta)=\frac{mH}{\hbar^2k_{F}cos(\Theta)}$, where $\Theta$ is the angle between the incident quasiparticle and the interface normal ($\hat{x}$). The $\delta$ function barrier takes into account the effect of any oxide barrier that may be present at the interface, as well as an effective barrier that might arise from the fermi velocity mismatch between the normal metal and the superconductor.\cite{1982PhRvB..25.4515B} We define the x direction as the axis perpendicular to the superconductor-normal interface (figure \ref{fig:gap}), and assume only the magnitude of the pair potential varies in the x direction. The angular dependence does not change as we go along x, since we expect the pair potential to be uniform in the y-z plane (perpendicular to the x direction). We note that a key assumption made here is that the inplane momentum is conserved upon tunneling, and no other orders emerge at the interface/proximity region. While this may not be realistic in interfaces between systems of different symmetry, it is a good first approximation. Furthermore, we anticipate the loss of in-plane momentum conservation to primarily broaden features, which could be included through adding lifetime effects to the model. However the subject of inelastic scattering in such structures is quite complex,\cite{belogolovskii2003phase,Pekola:2010ia} though we do not anticipate significant changes in our results. Ultimately, we assume a parabolic x dependence for the shape of the pair potential with $x_N$ and $x_S$ being the length of the induced superconducting region and the length of the reduced superconducting region respectively (figure \ref{fig:gap}). This approach has been shown to work well for s-wave superconductors, even when other functional forms of $\Delta(x)$ with gradual variations are used, so long as x$_N$, x$_S$, and the values of the induced and reduced gap ($\Delta_i$ and $\Delta_i$) are the same.\cite{PCvanSon:2011wx} We also note that this phenomenological model could be applied to an S/I/S' junction of two d-wave superconductors connected along the c-axis. Such a situation would require one to have $T_{c}>T>T_{c}'$, $x_{N}=\infty, x_{S}=0$, $\Delta_{r}=\Delta_{0} > \Delta_{i}$, however since we are interested in proximity junctions, we do not address this situation here. 

\section{Solving the Bogoliubov Equations}

We use the Bogoliubov equations to describe the quasiparticle excitations in a superconductor\cite{1982PhRvB..25.4515B,PCvanSon:2011wx}, determine the wavefunctions throughout the structure and ultimately to determine the probabilities of Andreev (A(E)) and normal reflection (B(E)) which govern the conductance. We note that in principle the problem could be solved analytically. In particular one would begin by considering only incoming electrons and outgoing holes (resulting from Andreev reflection) in the normal region. Then in the induced and reduced regions one would have pairs of electrons and holes, both incoming and outgoing, with their wavefunctions determined by the Bogoliubov equations with a parabolic $\Delta(x)$. Finally in the superconducting region only outgoing quasiparticles would need to appear. For $x<x_{N}$ one could use standard Bloch states, and for the $x>x_{S}$ the solutions found by Kashiwaya et al.\cite{Kashiwaya:1996wz} Ultimately one would need to apply the continuity conditions at each interface to determine the transmission and reflection probabilities and thus the conductance. However even in a purely s-wave case, such an approach is quite cumbersome, and for the d-wave case even more so. We also note that were such an approach to be undertaken, it would be best to ultimately solve the entire problem self-consistently to determine the gap. Since our goal is to investigate the role of various parameters in the tunneling spectra and provide further evidence for the proximity effect, we focus on a numerical solution described in detail below. 

 The elementary excitations of a superconductor are made from the creation of an electron with wavevector k and annihilation of an electron with wavevector -k (or creation of a hole with wavevector k). In this formalism, the excitation is represented by a two-element vector $\psi$\cite{1982PhRvB..25.4515B,PCvanSon:2011wx,Kashiwaya:1996wz}: $\psi(\vec{r}) = (f(\vec{r}), g(\vec{r}))$. For an anisotropic superconductor the Bogoliubov equations are written as\cite{Kashiwaya:1996wz}:
 \begin{eqnarray}\label{BdGe}
E f(r,\hat{k}) = [-\frac{{\hbar}^2{\nabla_{r}}^2}{2m} - \mu + V(\vec{r})]f(r,\hat{k}) + \Delta(r,\hat{k})g(r,\hat{k})\\ \label{BdGh}
E g(r,\hat{k}) = -[-\frac{{\hbar}^2{\nabla_{r}}^2}{2m} - \mu + V(\vec{r})]g(r,\hat{k}) + \Delta(r,\hat{k})f(r,\hat{k})
\end{eqnarray}
where $\hat{k} =  \vec{k}/|\vec{k}|$, $V(\vec{r})$, $\mu(\vec{r})$, and $\Delta(\vec{r})$, are the electrical, chemical, and pair potentials experienced by the quasi-particles. For simplicity, we assume the chemical potential is a constant across the interface and take the usual assumption that $H\delta(X)$ accounts for the interface (as described above). In the normal material, far from the interface, $\Delta= 0$ and as expected the equations \ref{BdGe} and \ref {BdGh} reduce to the Schrodinger equation for electrons(holes). To allow for a d-wave superconductor we assume $\Delta(\hat{k})$ has a momentum-dependence described by $\Delta(\hat{k}) = \Delta_{\infty}\cos{2\alpha}$, where $\alpha$ is defined as the angle between $\hat{k}$ and the $\hat{k}_y$ axis in the y,z plane (see Fig \ref{fig:gap}B). To simplify the numerical integration, we divide out this oscillation with a $e^{i\vec{k_F}\cdot\vec{r}}$ term and take the trial solutions for f and g: $f(\vec{r},\hat{k}) = u(\vec{r},\hat{k}) e^{i\vec{k}_{F}\cdot\vec{r}}$ \& $g(\vec{r},\hat{k}) = v(\vec{r},\hat{k}) e^{i\vec{k}_{F}\cdot\vec{r}}$. We obtain 
 \begin{eqnarray} \label{eq:BdG6}
-iv_{F}\hat{k}\cdot\bigtriangledown u(\vec{r},\hat{k}) = Eu(\vec{r},\hat{k})-\Delta(\vec{r},\hat{k})v(\vec{r},\hat{k})\\
iv_{F}\hat{k}\cdot\bigtriangledown v(\vec{r},\hat{k}) = Ev(\vec{r},\hat{k})-\Delta(\vec{r},\hat{k})u(\vec{r},\hat{k})\label{eq:BdG7}
\end{eqnarray}

Ultimately we are interested in calculating the probability of a quasiparticle incident on the interface traversing across, with no incoming quasi-particles on the superconducting side. For convenience it is easier to start with outgoing waves on the superconducting side, and integrating equations \ref{eq:BdG6} \& \ref{eq:BdG7} through the intermediate regions ($x_{N}< x < x_{S}$). Since we are working with an anisotropic superconductor, we have to account for the pair potentials difference for the electron ($k_{S}^+$) and hole ($k_{S}^{-}$) like quasiparticles: $\Delta_{\pm}=\Delta(\pm \hat{k}_{S}^{\pm})=|\Delta_{\pm}|exp(i\phi_{\pm})$, with $e^{i\phi_{\pm}}=\frac{\Delta_{\pm}}{|\Delta_{\pm}|}$ the phases of the pair potentials and $\Omega_{\pm}=\sqrt{E^{2}-|\Delta_{\pm}|^{2}}$. For a general x we expect the wavefunction to be of the form: $\psi_{j}(\vec{r},\hat{k})=(u_{aj}(\vec{r},\hat{k}), v_{aj}(\vec{r},\hat{k}))exp(-i\vec{k}_{S}^{-}\cdot\vec{r})+(v_{bj}(\vec{r},\hat{k}), u_{bj}(\vec{r},\hat{k}))exp(i\vec{k}_{S}^{+}\cdot\vec{r})$, where the subscripts a/b account for the left/right moving waves and j=(1,2) refers to solutions to equations \ref{eq:BdG6} and \ref{eq:BdG7}, respectively. In addition, due to the anisotropic nature, we will need to solve the equations or each value of the energy of the quasiparticle (as in BTK), but also for every incident ($\hat{k}\cdot \hat{x} = cos(\Theta)$) and azimuthal ($\alpha$) angle. Below we first describe how we solve for a given ($\Theta,\alpha$), then in the next section we show how to put this together to determine the conductance.  

In the superconductor ($x>x_{S}$) we only have the outgoing waves and thus the two independent solutions: 
\begin{eqnarray}
\label{eq:WFsup}
\psi^{S}_{1} = (u^{+}_{0}(E),e^{-i\phi_{+}} v^{+}_{0}(E)) e^{i\vec{k}_{S}^+\cdot\vec{r}} \\ \nonumber
\psi^{S}_{2} =(v^{-}_{0}(E), e^{-i\phi_{-}}u^{-}_{0}(E)) e^{-i\vec{k}_{S}^{-}\cdot\vec{r}}
\end{eqnarray}
where: 
\begin{eqnarray}
\label{eq:supinit}
u^{\pm}_{0}(E) = \sqrt{\frac{E+\Omega_{\pm}}{2E}} ~\&~ v^{\pm}_{0}(E) =\sqrt{\frac{E-\Omega_{\pm}}{2E}} 
\end{eqnarray}
Thus equations \ref{eq:WFsup} \& \ref{eq:supinit}  form the initial condition for each energy, and set of angles. In the intermediate region, ($-x_{N}<x<x_{S}$), the wavefunctions have the general form: 
\begin{eqnarray}
\psi^{l}_{j} = & (u^{+}_{j}(E,\vec{r}), e^{-i\phi_{+}(\vec{r})}v^{+}_{j}(E,\vec{r})) e^{i\vec{k}_{S}^+\cdot\vec{r}} \\ \nonumber
&+ (v^{-}_{j}(E,\vec{r}),  e^{-i\phi_{-}(\vec{r})}u^{-}_{j}(E,\vec{r})) e^{-i\vec{k}_{S}^-\cdot\vec{r}}
\end{eqnarray}
where $l=RS,IS$ for the reduced and induced superconducting regions respectively, ($u^{\pm}_{j}(E,\vec{r}),  v^{\pm}_{j}(E,\vec{r})$) are solutions to equations \ref{eq:BdG6} \& \ref{eq:BdG7}. At $x=x_{S}$, we set $\psi^{S}(x_{S})=\psi^{RS}(x_{S})$ and $\frac{d\psi^{S}}{dx}|_{x_{S}}=\frac{d\psi^{RS}}{dx}|_{x_{S}}$, neglecting the terms proportional to $\frac{du^{\pm}_{j}}{dx}$, $\frac{du^{\pm}_{j}}{dx}$, $\phi_{\pm}(x)$ and  $\vec{k}_{S}^{\pm}$. Next we use these conditions to numerically integrate equations \ref{eq:BdG6} \& \ref{eq:BdG7} from $x=x_{S}$ to $x=0$. At the boundary between the materials ($x=0$), we again match the boundary conditions, however accounting for the potential barrier of height (Z) (i.e. $\frac{d\psi^{RS}}{dx}|_{0}-\frac{d\psi^{IS}}{dx}|_{0}=2Z(\Theta)\psi^{RS}|_{0}$). These then become the initial conditions for the numerical integration of \ref{eq:BdG6} \& \ref{eq:BdG7} from the interface ($x=0$) to the normal region ($x=-x_{N}$). Finally we reach the normal region where the wavefunction has the form: 
\begin{eqnarray}
\psi^{N}_{j} = & (1,0) e^{i\vec{k}_{N}^+\cdot\vec{r}}+ a^{+}(E,\Theta,\alpha)(0,1) e^{-i\vec{k}_{N}^-\cdot\vec{r}}+ b^{+}(E,\Theta,\alpha)(1,0) e^{-i\vec{k}_{N}^+\cdot\vec{r}} \\ \nonumber
&+ (0,1) e^{-i\vec{k}_{N}^-\cdot\vec{r}}+ a^{-}(1,0)(E,\Theta,\alpha) e^{-i\vec{k}_{N}^+\cdot\vec{r}}+ b^{-}(E,\Theta,\alpha)(0,1) e^{-i\vec{k}_{N}^-\cdot\vec{r}}
\end{eqnarray}
where the coefficients $a^{\pm}(E,\Theta,\alpha),b^{\pm}(E,\Theta,\alpha)$ are the amplitudes of the Andreev and ordinarily reflected wave for an incident electron/hole. By applying the continuity conditions at $x=x_{N}$, we ultimately determine $a^{\pm},b^{\pm}$. 

\section{Calculation of the conductance spectra}
When an incoming electron is incident on a normal-superconductor interface, it can either ordinarily reflect at the interface, transmit through the interface with a wavevector on the same side of the fermi surface, transmit with a wavevector on the opposite side of the fermi surface, or Andreev-reflect.\cite{1982PhRvB..25.4515B} Andreev reflection involves two incoming electrons crossing the interface in low-barrier superconducting junctions and forming a Cooper pair, typically resulting in a doubling of conductance. Using the conservation of probability, the reflection and transmission coefficients can be related to one another. This leads to a simplified expression for the current at the normal-superconductor interface ($I_{NS}$), which depends solely on the probability of Andreev reflection ($A(E,\Theta,\alpha) = |a^{+}(E,\Theta,\alpha)|^2$) and the probability of ordinary reflection ($B(E,\Theta,\alpha)=|b^{+}(E,\Theta,\alpha)|^2$):\cite{1982PhRvB..25.4515B}.

 \begin{eqnarray}
I_{NS} \propto \int_{0}^{2\pi}\int_{-\pi/2}^{\pi/2}\int _{-\infty}^{\infty} [f(E-eV)-f(E)][1+A(E)-B(E)]dEd\Theta d\alpha
\end{eqnarray}

where f(E) is the fermi function (f(E) =$\frac{1}{1+e^{E/K_{B}T}}$), V is the applied voltage on the junction. Ultimately we are interested in the conductance across the interface ($\sigma_{NS}$), which can be determined from the current via: 
 \begin{eqnarray} \label{INS}
\sigma_{NS}(E,\Theta,\alpha) = \frac{dI}{dV}(E,\Theta,\alpha) = \frac{dI_{NS}}{dV}\\ \nonumber
\sigma (E) = \frac{\int d\alpha d\Theta \cos{\Theta} \sigma _{NS}(E,\Theta,\alpha)}{\int  d\alpha d\Theta \cos{\Theta} \sigma_N(E,\Theta,\alpha) }
\end{eqnarray}

The normalized conductance ($\sigma(E)$) is calculated numerically. Special care must be taken in dividing up the range for integration  as the resulting $a(E,\Theta,\alpha)~\&~b(E,\Theta,\alpha)$ oscillate strongly for certain values of $\Theta,\alpha$, depending on the initial parameters of $E,Z,x_{N},x_{S}$.

\section{Results and Discussion}
To check the validity of the calculation we first compared the resulting dI/dV to data from an optimally-doped Bi$_2$Sr$_2$CaCu$_2$O$_{8+x}$ (Bi-2212)/Bi$_2$Se$_3$ proximity device (described previously)\cite{2012NatCo...3E1056Z}. The red curve in figure \ref{fig:BS} shows a typical differential conductance from these proximity devices at 10 K, revealing a zero bias peak with a width much below the full gap (40 meV) and a second coherence peak also at a value below the Bi-2212 gap. These features were previously attributed to Andreev reflection arising from the induced superconductivity in the normal material, and the resulting reduced gap in the Bi-2212. The overall features are well reproduced by the calculations outlined here as shown by the red line where we used the parameters $\Delta_{induced} = 10 mV$, $\Delta_{reduced} = 24 mV$, $\Delta_{0} = 40 mV$,  Z= 0.5, x$_S$ = 2.6$\epsilon_0$, x$_N$ = 1.3$\epsilon_0$ (where $\epsilon_0$ = $\hbar v_F/(\pi \Delta_0)$ is the BCS coherence length). The mis-match in the fermi velocities are essentially included in the choice of Z, and only the ratio of the sizes of reduced and induced superconducting regions to the coherence length (x$_S$/$\epsilon_0$ and x$_N$/$\epsilon_0$) affect the calculation. 

Since our interface is perpendicular to the c-axis of the Bi-2212, the hole-like and the electron-like quasiparticles transmitted into the superconductor experience the same effective pair potentials, which have similar dependence on the azimuthal angle $\alpha$ in the ab-plane $\Delta+ = \Delta- = \Delta_0 cos(2\alpha)$. The calculated dI/dV is in good agreement with the experimental data (figure \ref{fig:BS}). We note that small discrepancies likely emerge from the fact that lifetime effects are not included in the calculation, nor are realistic band structures of the two materials, or strong-correlations. This would likely explain the apparent additional broadening in the spectra as well as the electron-hole asymmetry. Nonetheless the calculation clearly captures the qualitative features of the spectra and thus provides a good estimation of the spatial extent and shape of the induced superconducting gap. 

\begin{figure*}
\includegraphics[width=1\textwidth]{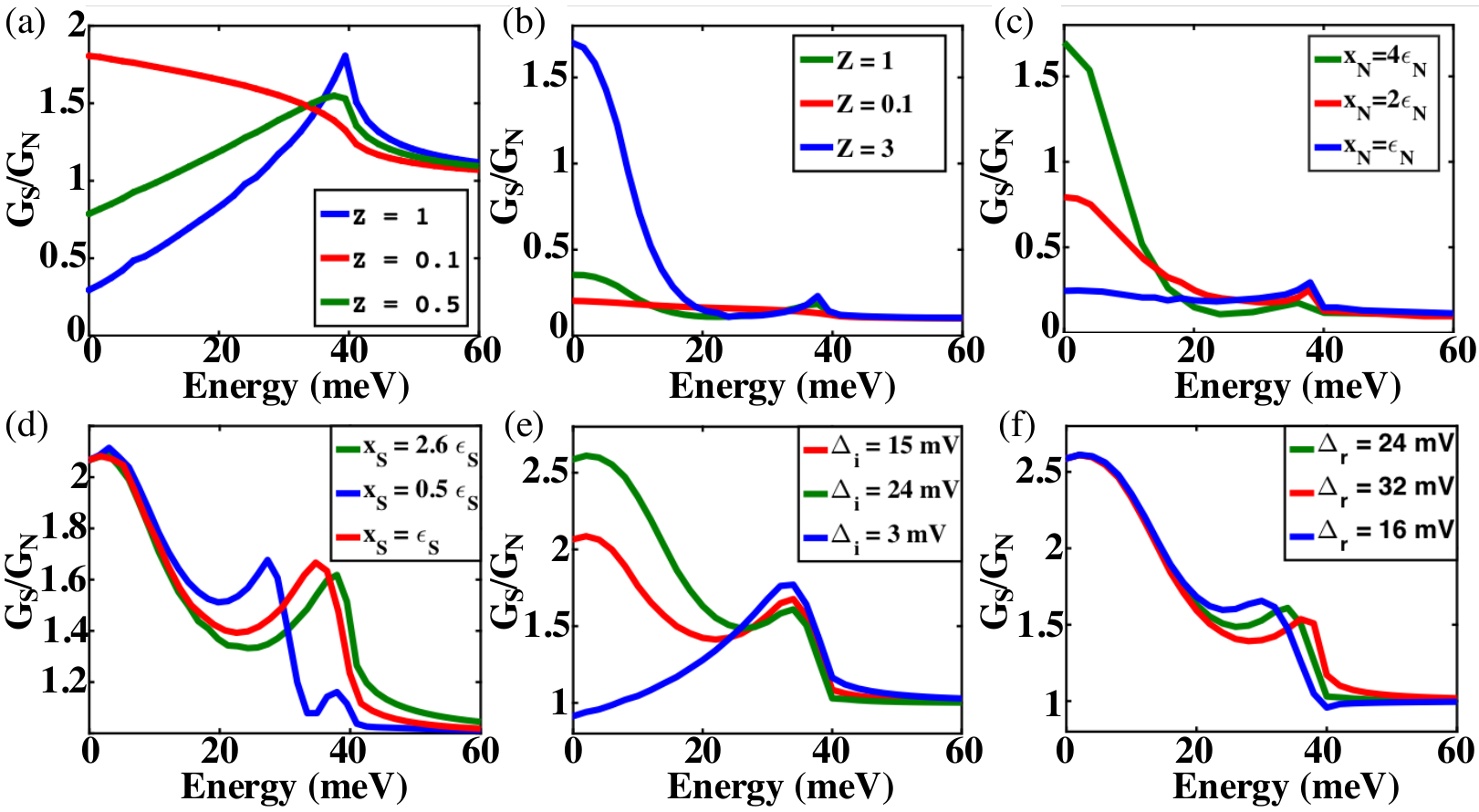}
\caption{\label{fig:fig3} (color online)(\textbf{A}) The dI/dV calculation for the parameters: $\Delta_{induced}$=0, $\Delta_{reduced}$ = $\Delta_0$ = 40 mV, x$_S$ = 2$\epsilon_0$, x$_N$ = 1.7$\epsilon_0$. All parameters are kept constant and the barrier height Z is varied. (\textbf{B}) The dI/dV calculation for the parameters: $\Delta_{induced}$=10 mV, $\Delta_{reduced}$ = 32 mV, $\Delta_0$ = 40 mV, x$_S$ = $\epsilon_0$, x$_N$ = 2$\epsilon_0$. All parameters are kept constant while z is varied. (\textbf{C}) The dI/dV calculation for the parameters: $\Delta_{induced}$=10 mV, $\Delta_{reduced}$ = 32 mV, $\Delta_0$ = 40 mV, x$_S$ = $\epsilon_0$, and Z = 3. All parameters are kept constant and x$_N$ is varied. (\textbf{D}) The dI/dV calculation for the parameters: $\Delta_{induced}$=10 mV, $\Delta_{reduced}$ = 24 mV, $\Delta_0$ = 40 mV, x$_N$ = 1.3$\epsilon_0$, Z = 0.6. All parameters are kept constant and the length of the reduced region, x$_S$, is varied from 0.5$\epsilon_0$ to $2.6\epsilon_0$. (\textbf{E}) The dI/dV calculation for the parameters: $\Delta_{reduced}$ = 36 mV, $\Delta_0$ = 60 mV, x$_S$ = x$_N$ = 1.3$\epsilon_0$, Z = 0.6. All parameters are kept constant and $\Delta_i$ is varied. (\textbf{F}) The dI/dV calculation for the parameters: $\Delta_{induced}$ = 16 mV, $\Delta_0$ = 60 mV, x$_S$ = x$_N$ = 1.3$\epsilon_0$, Z = 0.6. All parameters are kept constant and $\Delta_r$ is varied.}
\end{figure*}

In order to study the dependence of the conductance shape on the various parameters employed, we changed only one parameter while holding the others constant. The results are shown in figure \ref{fig:fig3}. First, we start with the simple case of no proximity effect (i.e. no leaking of the order parameter exists between the two materials). Therefore $\Delta_{reduced}$ = $\Delta_0$ = 40 mV and $\Delta_{induced}$ =0. Figure \ref{fig:fig3}a shows the calculated dI/dV spectra for x$_S$ = 2.6$\epsilon_0$ and x$_N$ = 1.7$\epsilon_0$ at various barrier heights. As expected, in the high barrier limit we obtain a coherence peak originating from the density of states of the superconductor (maximum at $\Delta_0$). Upon reducing the barrier, Andreev reflection begins to appear as an enhancement in zero bias conductance which eventually leads to a zero bias peak whose width reflects the full superconducting gap. Beyond confirming our naive expectations, these spectra are similar to the analytical calculation of \cite{Kashiwaya:1996wz} (figure 2 A in \cite{Kashiwaya:1996wz}). 

Next, we study the Z-dependence of the dI/dV spectra for a proximity-induced gap. figure \ref{fig:fig3} B shows the calculated dI/dV spectra for $\Delta_{induced}$=10 mV, $\Delta_{reduced}$ = 32 mV, $\Delta_0 = 40 mV$, x$_S$ = $\epsilon_0$, x$_N$ = 2$\epsilon_0$. The barrier height Z is varied from 0.3 to 1. As Z is increased, the peak structures are gradually enhanced, similar to the calculation of conductance spectra in high-barrier tunnel junctions for anisotropic superconductors by Kashiwaya et al. (figure 2 in \cite{Kashiwaya:1996wz}). However here we transfer from a regime where too low of a barrier leads to an Andreev reflection dominated by the original superconductor (i.e zero bias peak extending to the full gap), to a regime where a large barrier enables us to separate the proximity induced Andreev reflection (zero bias peak in blue curve extending to $\approx\Delta_{induced}$) from tunneling into the full gap of the superconductor (peak in the blue curve at $\Delta_{0}$). 

One may wonder why the reduced gap does not reveal any signatures in the spectra, and suggests we may not be capturing its effect on the tunneling. To investigate this further we varied the length of the induced(x$_N$) and reduced(x$_S$) superconducting regions. The results are shown in in figure \ref{fig:fig3} c and d, where all parameters are the same as in figure \ref{fig:fig3}b, however Z is fixed at 3. We find that the detailed spectra are quite sensitive to the choice of the induced and reduced regions, which is consistent with the sensitivity of tunneling to the shape of the wavefunction. For example in figure \ref{fig:fig3}c, we generally always observe the peak at the reduced gap, however the relative spectral weight of this and the zero bias peak are both quite sensitive to the spatial extent of the induced gap region. We believe this may result from a larger induced region allowing for a higher probability of Andreev tunneling into the induced gap. Interestingly, so far we have not observed a separate peak due to the full gap of Bi-2212 as well as the reduced gap. This is at odds with some of the data seen in junctions of Bi-2212 and Bi$_{2}$Te$_{3}$ or Bi$_{2}$Se$_{3}$.\cite{2012NatCo...3E1056Z}

An explanation this discrepancy emerges when the size of the reduced superconducting region is varied below a coherence length. In particular as shown in Fig \ref{fig:fig3}d, when x$_S$ is reduced below $\epsilon_0$, an additional peak shows up in the calculated spectra (the blue curve in figure \ref{fig:fig3}d), corresponding to the bulk superconducting gap. This peak appears due to an additional Andreev reflection process when the energy of the incoming quasiparticles is between the reduced and the bulk superconducting gap ($\Delta_{reduced} < E <\Delta_0$). At larger x$_S$, the variation of the pair potential at this region is very gradual and no peak is seen at $\Delta_0$. Perhaps not surprisingly, varying x$_{S}$ has a negligible, if any, effect on the width of the zero-bias peak as it is primarily governed by $\Delta_{induced}$ and Andreev tunneling into the proximity region. We note that the blue spectra in Fig \ref{fig:fig3}d corresponding to a reduced region half the width of the coherence length is consistent with most of our mechanically-bonded Bi-2212/normal material proximity devices\cite{2012NatCo...3E1056Z}. Such a small region over which the superconducting order is reduced is perhaps not surprising given the short coherence length in the Cuprates and the fact that the tunneling occurred along the c-axis. Finally, for a complete consistency check, we have also varied the size of $\Delta_{induced}$ or $\Delta_{reduced}$, while keeping all other parameters fixed. As shown in figure \ref{fig:fig3} e and f, the resulting spectra match our expectations. Specifically, lowering the induced gap results in a narrowing of the zero-bias peak, while lowing the reduced gap shifts the high energy peak in the spectrum to lower energies. 

\begin{figure*}
\includegraphics[width=1\textwidth]{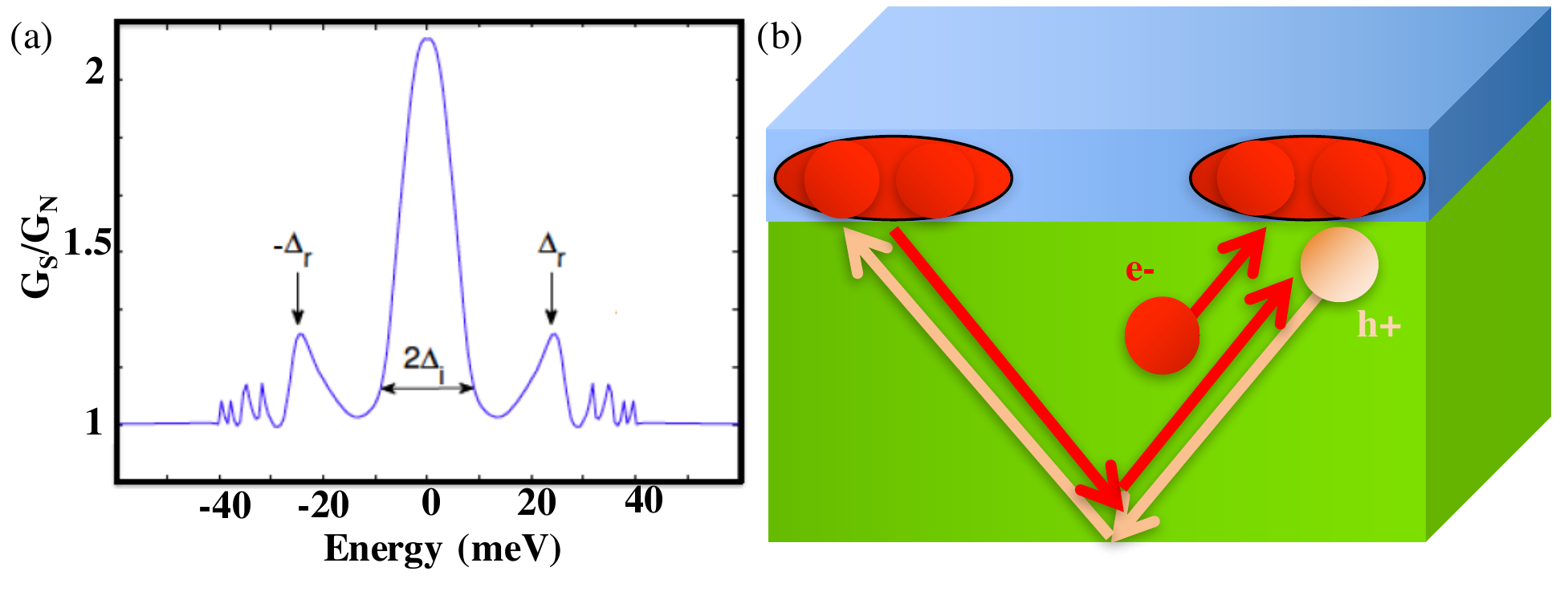}
\caption{\label{fig:fig4} (color online) (\textbf{A}) The differential conductance calculation for the proximity parameters $\Delta_{induced}$=12 mV, $\Delta_{reduced}$ = 28 mV, $\Delta_0 = 40 mV$, x$_S$ = 26$\epsilon_0$, x$_N$ = x$_S$/2, and Z = 0.6. Geometrical resonances are observed between the $\Delta_{reduced}$ and $\Delta_0$. (\textbf{B}) A schematic description of McMillan-Rowell oscillations.}
\end{figure*}

Interestingly we have found our approach can capture another subtle aspect of superconductor-normal metal heterostructures, namely coherent geometric resonances. These commonly observed features\cite{Wolf:1982uv,Deutscher:2005tm}  result from an interference of successive ordinary and Andreev reflections in thin layers, and are termed Tomasch and McMillan-Rowell oscillations, respectively\cite{Wolf:1982uv,Shkedy:2004kca,Chang:2004fl}. An example is shown in figure \ref{fig:fig4}a, where we plot the conductance for  $\Delta_{induced}$=12 mV, $\Delta_{reduced}$ = 28 mV, $\Delta_0 = 40 mV$, x$_S$ = 26$\epsilon_0$, x$_N$ = x$_S$/2, and Z=0.6. As discussed earlier, when the region over which the superconductivity is reduced is much larger then the coherence length, we expect to only observe features from the induced and reduced gaps. While both features are clearly seen, we also observe additional peaks between the reduced gap and the bulk superconducting gap values. To understand the origin of these features we recall the details of McMillan-Rowell and Tomasch oscillations. 

McMillan-Rowell oscillations occur due to interference effects in the normal material. Specifically, an electron in the normal material propagating towards the interface at an energy $E < \Delta_S$ is Andreev-reflected, sending a hole back. The reflected hole has a chance to ordinarily-reflect when it hits the normal material edge (x=d$_N$), return to the superconducting interface and Andreev-reflect again (figure \ref{fig:fig4}b). This leads to a series of reflections that interfere with each other, and appear as oscillatory features in the conductance spectra of this tunneling device. Similar to the described bound states in the normal layer, interference effects can occur in the potential well which is formed by the insulating barrier or the superconducting energy gap (Tomasch oscillations).\cite{PhysRevLett.74.2110} The oscillations we observe appear between $\Delta_{reduced}$ and $\Delta_0$ (figure \ref{fig:fig4}a) and thus are likely coming from an interference effect in the reduced superconducting region. We note that higher values of Z (e.g. see figure \ref{fig:fig3}b) did not produce any bound states that can emerge as seen in s-wave superconducting junctions.\cite{PhysRevLett.74.2110} However this is likely due to the subtle difference between d-wave and s-wave superconductors. Namely the node in the gap means bound-states are typically appearing at or very close to zero-energy.\cite{Deutscher:2005tm} Furthermore such states typically require a change in sign of the superconductor, and thus are seen in ab-plane tunneling into the [110] surface of the cuprates.\cite{2003PhyC..387..162G,Deutscher:2005tm} Indeed previous experiments on high-Tc junctions have only observed McMillan-Rowell and not Tomasch oscillations.\cite{1999PhRvB..60.9287N,2004PhyC..408..618C} However since we are studying c-axis heterostructures we do not expect to see such effects. We note that in superconductor/normal material heterostructures one may also expect to see resonances arising from Andreev Bound states, which are not accounted for in our approach. Such bound states are unlikely to be observed given the large innelastic scattering expected in such junctions. Nonetheless, the observation of geometrical resonances in our dI/dV calculations demonstrates the robustness of this approach. Our calculations not only capture the overall spectral features typically measured in unconventional superconducting proximity devices, but subtle bound states that can emerge as a result of interference effects. 

\section{Conclusion}
In this paper, we have used a phenomenological approach to compute the differential conductance spectra of a superconductor-normal proximity device for the case of an anisotropic superconductor by solving the Bogoliubov equations. We combine a successful approach in modeling s-wave proximity devices with a standard method for calculating the conductance of interfaces with d-wave superconductors. A parabolic, d-wave pair potential $\Delta (k,x)$ was assumed for the induced and reduced superconducting regions. A $\delta$ function was assumed to include any physical barrier as well any fermi velocity mismatch between the two materials. Under similar conditions the numerical calculation reproduces previous analytic solutions for d-wave superconductors. Furthermore in the proximity regime the calculated conductance spectra reveal a zero-bias peak whose width corresponds to the size of the induced superconducting gap in the normal material. Generally we also observe a peak corresponding to Andreev tunneling into the superconductor's reduced gap at the interface. Upon reducing the size of the reduced gap region to less then a coherence length, we also observe a peak at the full gap of the superconductor, consistent with the expectation that Andreev reflection now reaches the fully gapped region. The calculation is also able to reproduce geometric resonances that are well known to appear in thin heterostructures of normal metal-superconductors. Ultimately, we find that our results are consistent with experimental measurements of high-T$_c$ Bi-2212/Bi$_2$Se$_3$ devices and can be used to provide insights in interpreting tunneling into unconventional, superconducting proximity heterostructures. Nonetheless, this phenomenological model does not truly capture the real situation where the order parameter may be different in the induced region. Thus it would be helpful to extend this work by either self-consistently solving the BdG equations, properly modeling the wavefunction in the normal material (for example the Dirac states in topological materials), as well as accounting for a change in the phase of the order parameter across the boundary. 

\section{Acknowledgements}
The work at the University of Toronto was supported by the Natural Sciences and Engineering Research Council of Canada, the Canadian Foundation for Innovation, and the Ontario Ministry for Innovation. KSB acknowledge support from the National Science Foundation (grant DMR-1410846). The work at Brookhaven National Laboratory (BNL) was supported by DOE under Contract No. DE-AC02-98CH10886.
\vspace{1ex}

\bibliography{refe.bib}
\bibliographystyle{iopart-num}

\end{document}